\documentclass[twocolumn, pra]{revtex4-1}
\usepackage{amssymb}
\usepackage{amsmath}
\usepackage{epsfig}
\usepackage{color}
\usepackage{graphics, graphicx}
\usepackage{bbold}
\usepackage{psfrag}
\usepackage{mathcomp}
\usepackage{subfigure}
\usepackage{verbatim}
\usepackage{color}
\usepackage[colorlinks, citecolor=blue]{hyperref}

\begin{document}
\title{High-momentum distribution with subleading $k^{-3}$ tail in the odd-wave interacting one-dimensional Fermi gases} 
\author{Xiaoling Cui}
\email{xlcui@iphy.ac.cn}
\author{Huifang Dong}
\affiliation{Beijing National Laboratory for Condensed Matter Physics, Institute of Physics, Chinese Academy of Sciences, Beijing 100190, China}
\date{\today}
\begin{abstract}
We study the odd-wave interacting identical fermions in one-dimension with finite effective range. We show that to fully describe the high-momentum distribution $\rho(k)$ up to $k^{-4}$, one needs four parameters characterizing the properties when two particles {\it contact} with each other. Two parameters are related to the variation of energy with respect to the odd-wave scattering length and the effective range, respectively, determining the $k^{-2}$ tail and part of $k^{-4}$ tail in $\rho(k)$. The other two parameters are related to the center-of-mass motion of the system, respectively determining the $k^{-3}$ tail and the other part of $k^{-4}$ tail. We point out that the unusual $k^{-3}$ tail, which has not been discovered before in atomic systems, is an intrinsic component to complete the general form of $\rho(k)$ and also realistically detectable under certain experimental conditions. Various other universal relations are also derived in terms of those contact parameters, and finally the results are confirmed through the exact solution of a two-body problem.
\end{abstract}

\maketitle

\section{Introduction} 


Ultracold atoms exhibit universal properties in the strong coupling regime, where the system can be fully described by a few physical parameters and irrelevant to the short-range details of interaction potentials. In this regime, any perturbative approach fails to work, while a set of rigorous universal relations can be established to describe various microscopic and thermodynamic properties. These relations are connected by a key quantity called the {\it contact}, as first derived for a three-dimensional (3D) spin-1/2 Fermi gas near the s-wave Feshbach resonance\cite{Tan, Braaten, Zhang} and verified in experiments\cite{Jin_contact1,Jin_contact2,contact3}.  Later on the universal relations were generalized to other atomic systems, such as bosons\cite{Braaten1}, in low-dimension\cite{Werner, VZM, Zwerger,Valiente}, with higher partial-wave scatterings\cite{Ueda, Yu_p, Zhou, Cui, Yu_d} and with multiple scattering channels\cite{Zhou_new, Ueda_new, Qiran}. Very recently the contacts and universal properties of a 3D spin-polarized Fermi gas have been successfully explored near the p-wave Feshbach resonance\cite{Toronto}.

Among all the universal relations, the high-momentum distribution $\rho(k\rightarrow\infty)$ describes the microscopic property of the system and can be most easily detected in experiments by the time-of-flight technique. In general, the leading order of $\rho(k)$ in large $k$ limit is determined by the scattering channel of the system. For instance, the s-wave scattering gives rise to the leading $k^{-4}$ tail\cite{Tan, Braaten, Zhang,Werner, VZM, Zwerger,Valiente} while the p-wave gives the $k^{-2}$ tail\cite{Ueda, Yu_p, Zhou, Cui}, regardless of the dimension of the system. The sub-leading order, however, can be related to a number of effects, such as the effective range\cite{Braaten, Platter, Yu_p, Zhou}, the center-of-mass motion\cite{Yu_p, HuHui}, etc.  For instance, it has been found that these effects can contribute to the subheading $k^{-6}$ tail in s-wave\cite{Platter, HuHui} and $k^{-4}$ in p-wave\cite{Yu_p,Zhou}. Nevertheless, the odd powers of $k^{-1}$ have been exclusive in all previous studies, except the $k^{-5}$ tail found to be induced by Efimov physics\cite{Braaten1}.

In this work, we study the high-momentum distribution and universal relations of the odd-wave interacting Fermi gases in one-dimension. Our previous work based on the single-channel contact model has shown the interaction renormalization of this system and the leading high-momentum tail as $\rho(k)\sim C/k^2$\cite{Cui}. In this work, we extend our previous study to further include the finite range effect  through a two-channel model. 
We show that this treatment allows us to expand $\rho(k)$ as:
\begin{equation}
\rho(k)=\frac{C_l}{k^2}+\frac{C_{Q1}}{k^3}+\frac{2C_r+(3/4)C_{Q2}}{k^4}+o(k^{-5}). \label{k_tail}
\end{equation}
Therefore, to fully characterize the asymptotic behavior of  $\rho(k)$ up to $k^{-4}$, one needs four parameters characterizing the properties when two particles {\it contact} with each other. Two parameters ($C_l,\ C_r$) are related to the variation of energy with respect to the odd-wave scattering length and the effective range, respectively, determining the $k^{-2}$ tail and part of $k^{-4}$ tail in $\rho(k)$. The other two parameters ($C_{Q1},\ C_{Q2}$) are related to the center-of-mass motion of the system, respectively determining the $k^{-3}$ tail and the other part of $k^{-4}$ tail. We point out that the unusual $k^{-3}$ tail, which has not been discovered before in any atomic system, is an intrinsic component to complete the general form of $\rho(k)$. Its coefficient, $C_{Q1}$, can be related to the mean velocity of closed-channel dimers, and can be detected under certain realistic situations. We further derive various universal relations in terms of the contact parameters, and confirm our results  through the solution of a two-body problem. This work paves the way for studying the strongly odd-wave interacting Fermi gases in realistic quasi-1D geometry with finite effective range, and the results obtained also shed light on the general behavior of high-momentum distribution in cold atomic systems as well as the universal properties therein.


The rest of the paper is organized as follows. In section II we construct the two-channel model for identical fermions and establish the interaction renormalization. We derive the high-momentum distribution in section III and various other universal relations in section IV. These results are confirmed in section V by exactly solving a two-body problem. 
The last section VI is contributed to the discussion and summary of our results.

\section{Two-channel model and interaction renormalization} 

We start from the local Lagrangian (at coordinate $R$) of the identical fermions in 1D with odd-wave interaction:
\begin{eqnarray}
{\cal L}&=&\psi^{\dag}\left( i\partial_t+\frac{\partial_R^2}{2m} \right)\psi + d^{\dag}\left( i\partial_t-\nu_0 +\frac{\partial_R^2}{4m} \right)d + \nonumber\\
&&\frac{g}{2}\left( d^{\dag} [(i\partial_R\psi)\psi-\psi(i\partial_R\psi] +h.c. \right).
\end{eqnarray}
Here $\psi,\psi^{\dag}$ ($d,\ d^{\dag}$) are the field operators of atoms (dimers) in terms of the time $t$ and the coordinate $x$; $\nu_0$ is the bare detuning of the dimer; $g$ is the bare coupling between atoms and dimers. Accordingly, one can write down the many-body Hamiltonian as:
\begin{eqnarray}
{\cal H}&=&\sum_k \frac{k^2}{2m} \psi_k^{\dag}\psi_k + \sum_Q (\nu_0+ \frac{Q^2}{4m}) d_Q^{\dag}d_Q + \nonumber\\
&&\frac{g}{\sqrt{L}}\sum_{Q,k}\left(k d_Q^{\dag} \psi_{Q/2+k}\psi_{Q/2-k} +h.c. \right), \label{Hamil}
\end{eqnarray}
with $L$ the length of the system.

\begin{figure}[h]
\includegraphics[width=8.5cm]{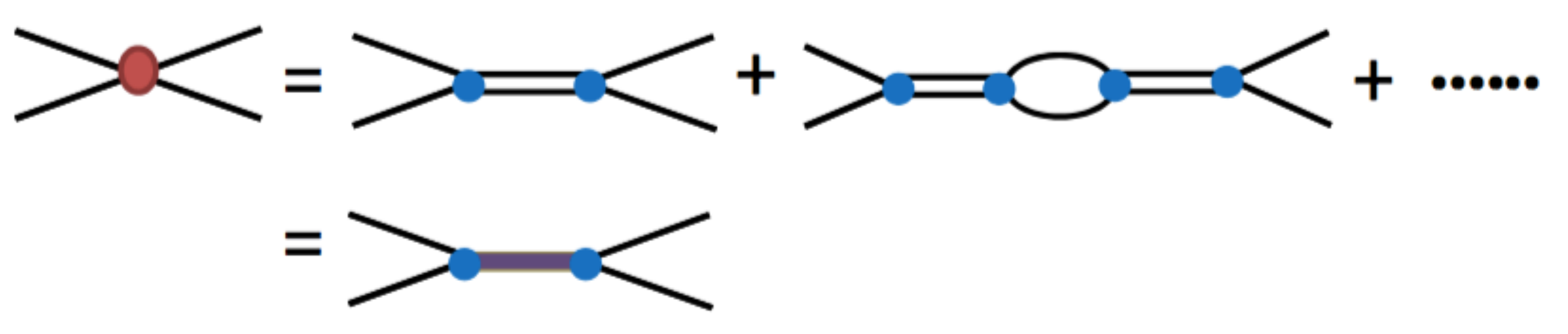}
\caption{(Color online). Diagrams for calculating the T-matrix (red solid circle). Here we consider the effective scattering between two atoms with incident momenta $(Q/2+k,Q/2-k)$ and outgoing momenta $(Q/2+k',Q/2-k')$.  The blue solid circle denotes the atom-dimer coupling vertex. The horizontal double lines represent the bare dimer propagator $D_0$, and the bold ones denote the renormalized dimer propagator $D$.  } \label{fig:schematic}
\end{figure}

Given the two-channel model, we first study the effective scattering between two atoms. Fig.1 shows the diagrams for two fermions scattering with total momentum $Q$ and total energy $E=Q^2/(4m)+k^2/m$ (the relative momenta of the incident and outgoing two-fermion are respectively $k$ and $k'$). The T-matrix can be obtained by summing up all relevant diagrams in Fig.1 (See Appendix A for more details on Feynman rules):
\begin{eqnarray}
-i\frac{T(E,Q)}{kk'}&=&D_0 \left(\frac{-2ig}{\sqrt{L}}\right)^2+D_0^2 \left(\frac{-2ig}{\sqrt{L}}\right)^4 \Pi+ ...
\nonumber\\
&\equiv&D \left(\frac{-2ig}{\sqrt{L}}\right)^2  \label{T}
\end{eqnarray}
here $D_0$ is the bare dimer propagator $D_0(E,Q)=i/(E-Q^2/(4m)-\nu_0)$, and $\Pi$ is polarization bubble:
\begin{equation}
\Pi(E,Q)=\frac{1}{2}\frac{iL}{2\pi}\int dq \frac{q^2}{E-Q^2/(4m)-q^2/m+i0^+} \label{Pi}
\end{equation}
Here the $\frac{1}{2}$ factor in front of above equation is to avoid the double counting of intermediate scattering states of identical fermions.  According to Eq.\ref{T}, the bold dimer propagator can be obtained via
\begin{equation}
D^{-1}(E,Q)=D_0^{-1}(E,Q)- \left(\frac{-2ig}{\sqrt{L}}\right)^2 \Pi(E,Q) ,
\end{equation}
and finally the T-matrix is expressed as:
\begin{equation}
\frac{k'k}{T(E,Q)}=\frac{mL}{4}  \left(\frac{1}{l_o}+r_o k^2+i k \right), \label{T-lp}
\end{equation}
where the odd-wave scattering length $l_o$ and the effective range $r_o$ are related to the bare scattering parameters in the two-channel model:
\begin{eqnarray}
\frac{m}{2l_o}&=&-\frac{\nu_0}{2g^2} + \frac{m\Lambda}{\pi} ;\label{lo}\\
r_o&=&\frac{1}{m^2g^2}. \label{ro}
\end{eqnarray}
Note that we have imposed a momentum cutoff $\Lambda$ in Eqs.(\ref{T},\ref{Pi}) to obtain the expression of $l_o$ in Eq.\ref{lo}. To compare with the interaction renormalization in single-channel model\cite{Cui}, we see that the bare atom-atom coupling $U$ there is replaced by $-2g^2/\nu_0$ in the present two-channel model, while the renormalization term ($\propto \Lambda$) is the same in both models. The main advantage of the two-channel framework here is that it naturally produces a finite effective range (Eq.\ref{ro}), to simulate the realistic situation when reducing the 3D p-wave interaction of identical fermions\cite{K40, K40_2, Li6_1, Li6_2} to the effective 1D odd-wave interaction by applying transverse confinements\cite{CIR_p_expe, CIR_p_1, CIR_p_2, CIR_p_3}.

\section{High-momentum distribution}

In this section, we derive the high-momentum distribution for identical fermions based on the two-channel model. Given the established interaction renormalization in last section, we can apply the quantum field method of operator-product-expansion(OPE)\cite{Wilson, Kadanoff}, which has been successfully applied to extracting universal relations in various interacting atomic systems\cite{Braaten, Zwerger, Cui, Yu_d}. 

We start from the expression of momentum distribution $\rho(k)$:
\begin{equation}
\rho(k)=\int dR \int dx e^{-ikx} \langle \Psi^{\dag}(R-\frac{x}{2}) \Psi(R+\frac{x}{2})\rangle.  \label{rho_k}
\end{equation}
Since the large-$k$ behavior of $\rho(k)$ is essentially determined by the one-body density matrix $\langle \Psi^{\dag}(R-\frac{x}{2}) \Psi(R+\frac{x}{2})\rangle$
at short distance $x\rightarrow 0$, we can utilize the OPE for expansion: 
\begin{equation}
\Psi^{\dag}(R-\frac{x}{2}) \Psi(R+\frac{x}{2})=\sum_n C_n(x) {\cal O}_n(R). \label{ope}
\end{equation}
Here the local operator ${\cal O}_n(R)$ can be constructed by quantum fields and their derivatives; the short-distance coefficient $C_n(x)$ can have non-analytic dependence on $x$, giving the power law tail of $\rho(k)$ according to Eq.(\ref{rho_k}).
In the following, we will extract a few leading non-analytical terms in $C_n(x)$ and identify their according local operators ${\cal O}_n(R)$.

As the OPE equation (\ref{ope}) is an operator equation, $C_n(x)$ can be determined by calculating the expectation value of each operator in (\ref{ope}) in the simplest two-body state $|Q/2+q,Q/2-q\rangle$. This state describes two colliding fermions with momenta $Q/2+q$ and $Q/2-q$ and with total energy $E=Q^2/(4m)+q^2/m$. 

\begin{widetext}

\begin{figure}[h]
\includegraphics[width=14cm]{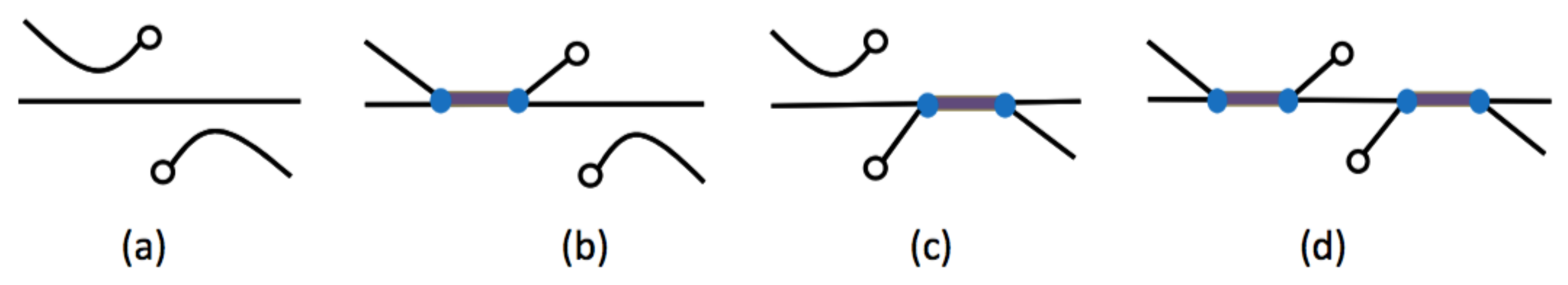}
\caption{(Color online). Diagrams for matrix elements of bi-local operator $\Psi^{\dag}(R-\frac{x}{2}) \Psi(R+\frac{x}{2})$ between the two-particle scattering states (See Appendix A for more details on Feynman rules). } \label{fig2}
\end{figure}

\begin{figure}[h]
\includegraphics[width=14cm]{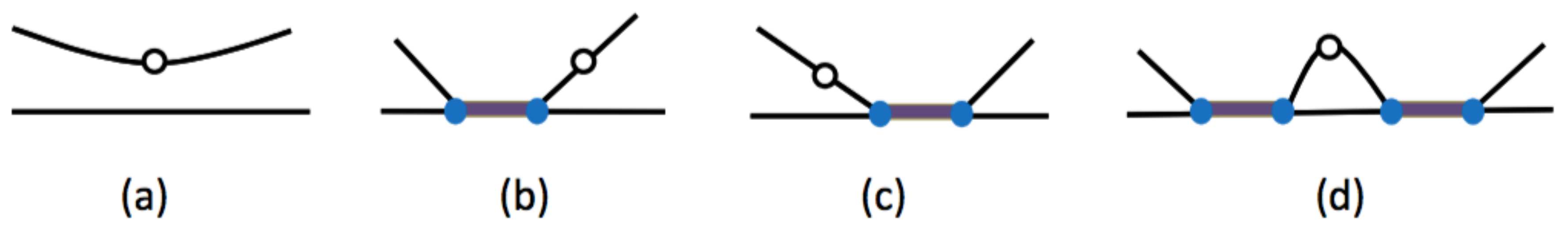}
\caption{(Color online). Diagrams for matrix elements of one-atom local operators, such as $\Psi^{\dag} \Psi(R)$ and its derivatives, between the two-particle scattering states (See Appendix A for more details on Feynman rules).} \label{fig4s}
\end{figure}

\end{widetext}

The left side of OPE equation (\ref{ope}) can produce four types of diagrams, as shown by Fig.2(a-d). According to the Feynman rules (see Appendix A), Fig.2(a) simply gives $g(x)\equiv (e^{i(Q/2+q)x}+e^{i(Q/2-q)x})/L$; Fig.2(b) and (c) both produce the same $x$-dependent $g(x)$, with an additional bare atomic propagator and a factor $q^2 (-2ig/\sqrt{L})^2 D(E,Q)/2$ from two atom-dimer scattering vertices and a bold dimer propagator; Fig.3(d), which involves an integral over the momenta of internal lines, gives: 
\begin{widetext}
\begin{eqnarray}
q^2 \left(\frac{-2ig}{\sqrt{L}}\right)^4 D^2(E,Q) i^3 \int \frac{dp dp_0}{(2\pi)^2} \frac{p^2 e^{ix(p+Q/2)}}{(p_0-(Q/2-p)^2/(2m)+i0^+)(E-p_0-(Q/2+p)^2/(2m)+i0^+)^2}=G(q; E,Q) f(x) \nonumber \\ \label{eq_d} 
\end{eqnarray}
\end{widetext}
with
\begin{eqnarray}
G(q; E,Q)&=&\frac{4m^2g^2}{L} q^2 \left(\frac{-2ig}{\sqrt{L}}\right)^2 D^2(E,Q); \label{G}\\
f(x)&=&\frac{i(1+i|q||x|)e^{i|q||x|}}{4|q|}e^{iQx/2}. \label{f}
\end{eqnarray}
We see that above $f$-function include a series of terms that are non-analytical at $x\rightarrow 0$:
\begin{equation}
-\frac{1}{2}|x|-\frac{iQ}{4}x|x| +(\frac{q^2}{6}+\frac{Q^2}{16})|x|^3 +o(|x|x^3). \label{f_non_analy}
\end{equation}
As discussed before, these terms will determine the large-$k$ behavior of $\rho(k)$.

Next we search for the local operators ${\cal O}_n(R)$ in the right side of OPE (\ref{ope}) to match the elements of $ \Psi^{\dag}(R-\frac{x}{2}) \Psi(R+\frac{x}{2})$ as produced by the diagrams in Fig.2(a-d). First, we note that Fig.2(a,b,c) all produce analytical function of $x$, thus one can simply find the local operators by Taylor expanding these functions  in terms of $x$. Take the simplest function $e^{ikx}$ for instance, as $e^{ikx}=\sum_n (ik)^n x^n/n! \ (n=0,1,...)$, so its corresponding local operators (connecting an incoming and an outgoing atomic line with the same momentum $k$) are simply the one-body operators ${\cal O}_n(R)=\psi^{\dag}(\overrightarrow{\partial_R})^n\psi(R)$\cite{footnote_localoperator}, and the coefficients are $C_n(x)=x^n/n!$. Following that, one can show that the diagrams in Fig.3(a,b,c) respectively match the expansions in power of $x$ of the corresponding diagrams in Fig.2(a,b,c). The diagrams in Fig.3(d) can match  the analytical functions of $x$ produced by the diagrams in Fig.2(d). However, Fig.2(d) also produces non-analytical terms as in the f-function in Eqs.(\ref{f},\ref{f_non_analy}),  which cannot be matched by any diagram in Fig.3(a-d). These non-analytical terms  must correspond to other ${\cal O}_n(R)$  other than the one-atom local operators.

\begin{figure}[h]
\includegraphics[width=6cm]{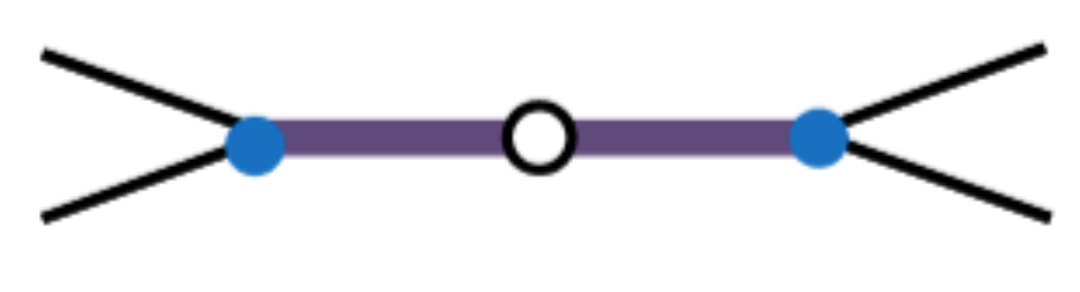}
\caption{(Color online). Diagrams for matrix elements of dimer local operators, such as $d^{\dag} d(R)$ and its derivatives, between the two-particle scattering states.} \label{fig4s}
\end{figure}


In order to match the non-analytical terms produced by Fig.2(d), we have to find the local operator ${\cal O}_n(R)$ whose expectation value  under  $|Q/2+q,Q/2-q\rangle$ is proportional to $G(q,E,Q)$. It turns out to be the dimer local operator $d^{\dag}(R)d(R)$ or its derivatives, as shown in Fig.4. These operators and their according matrix elements are:
\begin{eqnarray}
&&\langle C_l(R) \rangle\equiv 4m^2g^2\langle d^{\dag}(R)d(R) \rangle =G(q;E,Q); \label{C_l}\\
&&\langle C_r(R) \rangle\equiv 4m^3g^2\langle d^{\dag}(R) (i\partial_t+\frac{\partial_R^2}{4m}) d(R) \rangle=G(q;E,Q)q^2; \nonumber\\
\label{C_r}\\
&&\langle C_{Q1}(R) \rangle\equiv 4m^2g^2\langle d^{\dag}(R) (-i\partial_R) d(R) \rangle=G(q;E,Q)Q; \label{C_Q1}\\
&&\langle C_{Q2}(R) \rangle\equiv 4m^2g^2\langle d^{\dag}(R) (-\partial_R^2) d(R) \rangle=G(q;E,Q)Q^2. \label{C_Q2}
\end{eqnarray}
Thus we have the following terms in the right side of OPE equation (\ref{ope}), 
\begin{equation}
-\frac{|x|}{2}\langle C_l(R) \rangle-\frac{ix|x|}{4}\langle C_{Q1}(R) \rangle +\frac{|x|^3}{6}\langle C_r(R) \rangle+\frac{|x|^3}{16}\langle C_{Q2}(R)\rangle,
\end{equation}
to match the  left side of (\ref{ope}). In combination with Eq. \ref{rho_k}, we obtain the large momentum distribution as Eq.(\ref{k_tail}), where the four parameters are defined as
\begin{equation}
C_{\xi}=\int dR \langle C_{\xi}(R) \rangle, \label{C_xi}
\end{equation}
here $\xi$ denotes the subscript $l, r, Q1$ or $Q2$. 

It is worth to point out that each $C_{\xi}$, defined in Eqs.(\ref{C_l}-\ref{C_Q2}) through the local operator of the dimer field, describes a particular property when two particles {\it contact} with each other. 
Specifically, $C_{l}$ is referred to as the contact density; $C_{Q1}$ is the contact current\cite{Nishida}; $C_r$ and $C_{Q2}$ are the combinations of contact density and momentum variance.  
Moreover, they can also be classified straightforwardly if the relative and center-of-mass motions of interacting particles are fully decoupled: $C_l$  and $C_r$ are only associated with relative motion, while $C_{Q1}$ and $C_{Q2}$ are also related to the center-of-mass motion. This can be seen more clearly from the exact solution of two-body problem as shown in section V. However, for a general case when the relative and center-of-mass motions coupled with each other, all the four quantities $C_{\xi}$ will be sensitively dependent on both of the two motions.

Following above procedures, we can further extend the $\rho(k)$ expansion to even higher orders of $k^{-1}$. It can be expected that all the odd powers of $|x|$, i.e., $|x|^{2n-1}$, in f-function will produce the $k^{-2n}$ tail in $\rho(k)$, while the $x|x|^{(2n-1)}$ terms in f-function will produce the $k^{-(2n+1)}$ tail in $\rho(k)$. These odd-power tails as $k^{-(2n+1)}$ have not be found in the large momentum distribution of atomic systems studied before, except the $k^{-5}$ tail in identical bosons due to a different mechanism of Efimov physics\cite{Braaten1}. Nevertheless, we show here that these odd-power terms can be caused by particular types of non-analyticity in the one-body density matrix at short distance, thus should all be included to complete the general form of $\rho(k)$ in the large $k$ limit. Possible detection of these odd power terms in $\rho(k)$ will be discussed later. 


\section{Universal relations}

In this section, we derive various other universal relations for identical fermions.

{\it (i) Adiabatic relation.} Using the renormalization equations (\ref{lo},\ref{ro}) and the Feynman-Hellman theorem, we can obtain the first derivative of total energy, $E$, with respect to the two scattering parameters:
\begin{eqnarray}
\frac{\partial E}{\partial (-1/l_o)} &=& \langle \frac{\partial {\cal H}}{\partial \nu_0} \rangle \frac{\partial \nu_0}{\partial (-1/l_o)} \nonumber\\
&=&mg^2\int dR \langle d^{\dag}(R)d(R) \rangle ; \label{adiab_1}\\
\frac{\partial E}{\partial r_o} &=& \langle \frac{\partial {\cal H}}{\partial g} \rangle \frac{\partial g}{\partial r_o} + \langle \frac{\partial {\cal H}}{\partial \nu_0} \rangle \frac{\partial \nu_0}{\partial r_o}\nonumber\\
&=&-m^2g^2\int dR \langle d^{\dag}(R) (i\partial_t+\frac{\partial_R^2}{4m}) d(R)\rangle  \label{adiab_2}
\end{eqnarray}
Note that in getting Eq.(\ref{adiab_2}) we have used the equation of motion for the dimer field. Further according to the definitions of $C_{\xi}$ in Eqs.(\ref{C_l},\ref{C_r},\ref{C_xi}), we obtain the adiabatic relations
\begin{eqnarray}
\frac{\partial E}{\partial (-1/l_o)} &=& \frac{C_l}{4m} ; \label{adia1}\\
\frac{\partial E}{\partial r_o} &=& -\frac{C_r}{4m} . \label{adia2}
\end{eqnarray}

{\it (ii) Energy relation.} By utilizing the equation of motion for dimer fields to simplify the  total energy $E=\langle {\cal H}\rangle$, 
we can write the energy as:
\begin{equation}
E=\int \frac{dk}{2\pi} \frac{k^2}{2m} \rho(k)-\frac{\nu_0}{g^2} \frac{C_l}{4m^2}+\frac{C_{Q2}}{16m^3g^2}+ \frac{C_r}{2m^3g^2}  
\end{equation}
Further using the renormalization equations (\ref{lo},\ref{ro}), we arrive at
\begin{equation}
E=\int \frac{dk}{2\pi} \frac{k^2}{2m} \left( \rho(k) - \frac{C_l}{k^2}\right) +\frac{C_l}{4ml_o}+\frac{r_o}{m}\left( \frac{C_{Q2}}{16}+ \frac{C_r}{2}\right).  
\label{energy}
\end{equation}
One can see that except for $C_{Q1}$, the other three parameters $C_l,\ C_r,\ C_{Q2}$ all explicitly show in the expression of $E$.  In the zero range limit $r_o\rightarrow 0$, Eq.\ref{energy} reproduces the result obtained within a single-channel  model\cite{Cui}.  Note that in the presence of a trapping potential $V_T$, an additional term, $\langle V_T\rangle$, should be added to Eq.\ref{energy}.

{\it (iii) Virial theorem.} With a harmonic trapping potential $V_T=\sum_i m\omega^2 x_i^2/2$, the Virial theorem can be obtained via dimensional analysis\cite{Tan, Braaten, Zhang, Zwerger}, which requires $(\omega \partial/\partial \omega -  l_o/2 \partial/\partial l_o-r_o/2 \partial/\partial r_o)E=E$. Using the  Feynman-Hellman theorem together with the adiabatic relations (\ref{adia1},\ref{adia2}),  we obtain
\begin{equation}
E = 2 \langle V_T\rangle-\frac{C}{8m l_o}+r_o\frac{C_r}{8m}.  \label{virial}
\end{equation}

{\it (iv) Pressure relation.} In a homogeneous system, the pressure relation can be obtained by applying dimensional analysis\cite{Tan, Braaten, Zhang, Zwerger} to the free energy density ${\cal F}=F/L$, which requires $(T \partial/\partial T +n/2 \partial/\partial n - l_o/2 \partial/\partial l_o-r_o/2 \partial/\partial r_o){\cal F}=3/2 {\cal F}$. Using the adiabatic relations (\ref{adia1},\ref{adia2}), the pressure density ${\cal P}$ can then be related to the energy density ${\cal E}$ as:
\begin{equation}
{\cal P}=2{\cal E}+\frac{{\cal C}_l}{4ml_o} - r_o\frac{{\cal C}_r}{4m}, \label{pressure}
\end{equation}
where ${\cal C}_{\xi}\equiv C_{\xi}/L\ (\xi=l,r) $.


\section{Two-body check}

It is useful to check the high-momentum distribution and universal relations through the simple yet insightful two-body problem. For two atoms and a dimer scattering in a homogenous system, we can write down an ansatz wave-function: 
\begin{equation}
|\psi\rangle=\left( \Phi d^{\dag}_Q + \sum_{k>0}  \phi_k \psi^{\dag}_{Q/2+k}\psi^{\dag}_{Q/2-k}\right) |{\rm Vac}\rangle
\end{equation}
Here we have considered a finite total momentum $Q$, a conserved quantity in this problem. By imposing the Schrodinger equation ${\cal H}|\psi\rangle=E|\psi\rangle$, we can get two equations in terms of the coefficients $\Phi$ and $\phi_k$, which finally result in a single equation for the binding energy $E_b\equiv E-Q^2/(4m)$:
\begin{equation}
\frac{E_b-\nu_0}{2g^2}=\int \frac{dk}{2\pi}\frac{k^2}{E_b-k^2/m}.
\end{equation}
Given the scattering parameters defined in (\ref{lo},\ref{ro}), the above equation is equivalent to
\begin{equation}
\frac{1}{l_o}-r_o\kappa^2-\kappa=0, \label{kappa}
\end{equation}
where $\kappa=\sqrt{-mE_b}$. Note that Eq.\ref{kappa} corresponds to the divergence of T-matrix in Eq.\ref{T-lp} with momentum $k$ replaced by $i\kappa$. 

Given the energy solution in Eq.\ref{kappa}, the momentum distribution of atoms can be obtained as:
\begin{eqnarray}
\rho(k)&=&\frac{|\phi_{k-Q/2}|^2\theta(k-Q/2)+|\phi_{Q/2-k}|^2\theta(Q/2-k)}{\sum_{k>0}|\phi_k|^2+\Phi^2}    \nonumber\\
&=& \frac{8\kappa}{1+2r_o\kappa} \frac{(k-Q/2)^2}{(k-Q/2)^2+\kappa^2} , \label{rho_2-body}
\end{eqnarray}
In the large $k$ limit, we get
\begin{equation}
\rho(k)= \frac{8\kappa}{1+2r_o\kappa} \left( \frac{1}{k^2}+\frac{Q}{k^3}+\frac{-2\kappa^2 +(3/4)Q^2}{k^4}+o(k^{-5}) \right)
\end{equation}
We see that this asymptotic form fits Eq.\ref{k_tail} well. The coefficients of all expansion terms are also checked to be consistent with the parameters defined in Eqs.(\ref{C_l},\ref{C_r},\ref{C_Q1},\ref{C_Q2},\ref{C_xi}), which under the two-body ansatz read\begin{eqnarray}
C_l&=&\frac{8\kappa}{1+2r_o\kappa};\ \ C_r=-\frac{8\kappa^3}{1+2r_o\kappa} ;\nonumber\\
C_{Q1}&=& \frac{8\kappa Q}{1+2r_o\kappa}; \ \ C_{Q2}=\frac{8\kappa Q^2}{1+2r_o\kappa}. \label{C_2-body}
\end{eqnarray}
One thus sees that in the two-body system where the relative and center-of-mass motions can be fully separated, $C_{Q1},\ C_{Q2}$ are simply proportional to $Q$ and $Q^2$, as already suggested by Eqs.(\ref{C_Q1},\ref{C_Q2}); while $C_l, \ C_r$ are solely determined by the relative motion. 

Similarly, one can verify straightforwardly the other universal relations expressed in Eqs.(\ref{adia1},\ref{adia2},\ref{energy},\ref{pressure}), by using the binding energy solution in Eq.\ref{kappa} and the parameters in Eq.\ref{C_2-body}. Take the adiabatic relations (\ref{adia1},\ref{adia2}) for instance, the left sides of (\ref{adia1}) and (\ref{adia2}) can be converted to $(2\kappa/m) \partial\kappa/\partial (1/l_o)$ and $-(2\kappa/m) \partial\kappa/\partial r_o$ respectively, while the derivatives $\partial\kappa/\partial (1/l_o)$ and $\partial\kappa/\partial r_o$ can be easily extracted from Eq.\ref{kappa}. Given all the known information of $\kappa,\ \rho(k)$ and $C_{\xi}$ from the two-body solution, one can verify all universal relations expressed in section IV.

\section{Discussion and Summary}

We have shown in this work that the odd-power terms as $k^{-(2n+1)}$ in the large momentum distribution $\rho(k)$ can originate from a particular set of non-analyticity in the one-body density matrix at short-distance. As a result, these terms should be included as intrinsic components of $\rho(k)$. Our results would shed light on the general form of high-momentum distribution in cold atomic systems studied previously and in future.

The $k^{-3}$ tail and the associated contact current $C_{Q1}$ revealed in this work cannot show up in a homogeneous 1D gas of identical fermions\cite{Bender}; instead, their existence requires additional conditions. 
Since $C_{Q1}$ can be expressed through the mean velocity of dimers $\bar{v}_d\equiv (1/m)\int dR \langle d^{\dag}(R)(-i\partial_R)d(R)\rangle$:
\begin{equation}
C_{Q1}=\frac{4m}{r_o} \bar{v}_d,
\end{equation}
we list below several promising systems to support finite $C_{Q1}$ or $\bar{v}_d$. 
First, they can appear in non-equilibrium system. For instance, a finite $\bar{v}_d$ can be developed dynamically by applying an external driving force such as the magnetic field gradient. Secondly, they can be probed in equilibrium system with broken time-reversal symmetry (TRS).   This can be more easily realized in spin-1/2 systems, where the p-wave interaction exists between different atomic isotopes (such as the two-species Li6 fermions\cite{Li6}).
The broken TRS can be induced by spin imbalance or by spin-orbit coupling as realized in current cold atoms experiments\cite{soc_review}. These systems have been found to support finite-momentum two-body bound state and pairing superfluidity when an attractive s-wave interaction is introduced between different species\cite{fflo, zhouprl}. We cannot find any reason that a p-wave interaction can change qualitatively the conclusion. Thirdly, they can exist in systems where the two-body interaction relies on the center-of-mass momentum of colliding particles\cite{PengZhang}.
In this case, the ground state of the system can host a finite $C_{Q1}$ that is purely driven by the interaction effect. In future, it is interesting to explore more accurately the  contact effects in these systems. 

In practice, $C_{Q1}$ can be extracted  from the leading order of the subtracted momentum distribution $\rho(k)-\rho(-k)=2C_{Q1}/k^3+o(k^{-5})$\cite{footnote}. 
We note that the other three parameters, $C_l,\ C_r$ and $C_{Q2}$, can be detected from various other universal relations in Eqs.(\ref{adia1},\ref{adia2},\ref{energy},\ref{virial},\ref{pressure}).

In summary, we have studied the high-momentum distribution and universal relations of odd-wave interacting identical fermions in one-dimension with finite effective range. In particular, we pointed out that the odd power terms as $k^{-(2n+1)}$ should be included in order to complete the general form of $\rho(k)$ in the large $k$ limit. Our results can be straightforwardly generalized to other odd-wave interacting 1D systems with spin degree of freedom, such as the fermion-fermion, fermion-boson, and boson-boson mixtures, which can be realized in view of the existent p-wave Feshbach resonance in these systems\cite{Li6_1,Li6_2,K40_2, FR_review}.



\bigskip

{\bf Acknowledgement.}
The work is supported by the National Natural Science Foundation of China (No.11374177, No. 11534014), the National Key Research and Development Program of China (2016YFA0300603) and the programs of Chinese Academy of Sciences.

\bigskip

\appendix

\section{Feynman rules}

We present the Feynman rules for evaluating the diagrams in OPE method. 

(1) {\it Two-momenta and loop integral.} Each external line is assigned with two-momenta ($p_0,p$), with $p_0$ the energy and $p$ the momentum. The two-momenta of incoming and outgoing lines are constrained by the conservation of total momentum and total energy. If there exists two-momenta  ($p_0,p$) independent on the two-momenta of incoming and outgoing lines, they should be integrated over using $L/(2\pi)^2\int dp dp_0 $.

(2) {\it Propagators.}  Each internal line with two-momenta ($p_0,p$) is assigned a propagator factor $i/(p_0-p^2/(2m)+i0^+)$ for atoms and $i/(p_0-\nu_0-p^2/(4m)+i0^+)$ for dimers. 

(3) {\it Atom-dimer scattering vertex.} Each scattering vertex between two identical fermions and a dimer is assigned a factor $-2ig/{\sqrt{L}}$ (note that the factor of 2 is absent if the two atoms are distinguishable).

\begin{figure}[h]
\includegraphics[width=9cm]{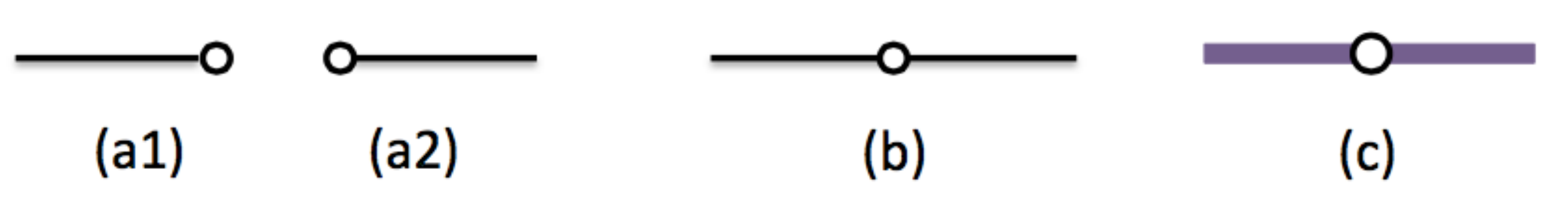}
\caption{(Color online). Diagrams of local operators: (a1) $\psi(R)$, (a2) $\psi^{\dag}(R)$; (b) one-atom local operators, such as $\psi^{\dag}(R)\psi(R)$ and its derivatives; (c) dimer local operators, such as $d^{\dag}(R)d(R)$ and its derivatives.} \label{fig2s}
\end{figure}

(4) {\it Operator vertices.}

Two basic local operators $\psi(R)$ and $\psi^{\dag}(R)$ are diagrammatically shown in Fig.5(a1) and (a2). $\psi(R)$ (or $\psi^{\dag}(R)$) is denoted by an open dot with an atom line ending (or starting) at the dot. If $\psi(R)$ (or $\psi^{\dag}(R)$) is linked to an atom line with momentum $k$, it will produce $e^{ikR}/\sqrt{L}$ (or $e^{-ikR}/\sqrt{L}$).

The one-atom local operators, which annihilate an atom and create an atom at the same site, are diagrammatically shown in Fig.5(b). These operators include $\psi^{\dag}(R)\psi(R)$ and its derivatives such as $\psi^{\dag}(R)({\partial_R})^m\psi(R)$ ($m$ is an integer), etc. 

The dimer local operators, which annihilate a dimer and create a dimer at the same site, are diagrammatically shown in Fig.5(c). These operators include $d^{\dag}(R)d(R)$ and its derivatives such as $d^{\dag}(R)({\partial_R})^m d(R)$ ($m$ is an integer), etc.


\begin{thebibliography}{99}
\bibitem{footnote_localoperator} For the bi-local operator which connects different incoming and outgoing momenta, i.e., $\langle p|\Psi^{\dag}(R-\frac{x}{2}) \Psi(R+\frac{x}{2})|p'\rangle=e^{iR(p-p')}e^{ix(p+p')/2}$, the Taylor expansion of $x$-dependent function is more involved. The corresponding local operator in this case can be written as $\psi^{\dag}(R)(\overleftarrow{\partial_R})^m(\overrightarrow{\partial_R})^n\psi(R)$ in general.
\bibitem{footnote} Note that in realistic quasi-1D system, the large momentum refers to the region $k_F\ll k \ll 1/a_{\perp}$, where $k_F$ is the Fermi momentum of the system and $a_{\perp}$ is the characteristic length of the transverse confinement potential. This condition can be achieved by preparing a small cluster of fermions or imposing deep transverse confinement such that $k_Fa_{\perp}\ll1$.
\end{thebibliography}
\end{document}